\documentclass[preprint,review,12pt]{elsarticle}

\usepackage{amssymb}

\usepackage{hyperref}
\usepackage{amsmath}
\usepackage{cleveref}
\usepackage{booktabs}
\usepackage{caption,setspace}
\usepackage{tabularx}
\usepackage[export]{adjustbox}
\usepackage{tikz}
\usepackage{geometry}
\usepackage{wrapfig} 
\usepackage{flushend} 
\usepackage{algorithm}
\usepackage{algpseudocode}
\usepackage{xcolor}
\usepackage{multirow}

\hypersetup{pdfauthor={Name}}

\makeatletter
\newcommand{\ssymbol}[1]{^{\@fnsymbol{#1}}}
\makeatother

\usepackage{array}
\newcolumntype{P}[1]{>{\centering\arraybackslash}p{#1}}

\newcommand{\ts}[1]{\textsuperscript{#1}}
\newcommand{\tb}[1]{\textbf{#1}}

\AtBeginDocument{
    \crefformat{equation}{#2Eq.~#1#3}
    \crefformat{figure}{#2Fig.~#1#3}
    \crefformat{table}{#2Tab.~#1#3}
}

\journal{ArXiv}  

\begin{document}
\begin{sloppypar}

\begin{frontmatter}

\title{UNGT: Ultrasound Nasogastric Tube Dataset for Medical Image Analysis}


\author[a]{Zhaoshan Liu\corref{cor1}}
\ead{e0575844@u.nus.edu}

\author[b]{Chau Hung Lee\corref{cor1}}
\ead{chau_hung_lee@ttsh.com.sg}

\author[c]{Qiujie Lv}
\ead{lvqiujie@zzu.edu.cn} 

\author[b]{Nicole Kessa Wee}
\ead{nicolekessa_wee@ttsh.com.sg}

\author[a]{Lei Shen\corref{cor2}}
\ead{mpeshel@nus.edu.sg}

\cortext[cor1]{Equal contribution}

\cortext[cor2]{Corresponding author}

\address[a]{Department of Mechanical Engineering, National University of Singapore, 9 Engineering Drive 1, Singapore, 117575, Singapore}

\address[b]{Department of Radiology, Tan Tock Seng Hospital, 11 Jalan Tan Tock Seng, Singapore, 308433, Singapore}

\address[c]{School of Computer and Artificial Intelligence, Zhengzhou University, 100 Science Avenue, Zhengzhou, 450001, China}


\begin{abstract}
We develop a novel ultrasound nasogastric tube (UNGT) dataset to address the lack of public nasogastric tube datasets. The UNGT dataset includes 493 images gathered from 110 patients with an average image resolution of approximately 879 $\times$ 583. Four structures, encompassing the liver, stomach, tube, and pancreas, are precisely annotated. Besides, we propose a semi-supervised adaptive-weighting aggregation medical segmenter to address data limitation and imbalance concurrently. The introduced adaptive weighting approach tackles the severe unbalanced challenge by regulating the loss across varying categories as training proceeds. The presented multiscale attention aggregation block bolsters the feature representation by integrating local and global contextual information. With these, the proposed AAMS can emphasize sparse or small structures and feature enhanced representation ability. We perform extensive segmentation experiments on our UNGT dataset, and the results show that AAMS outperforms existing state-of-the-art approaches to varying extents. In addition, we conduct comprehensive classification experiments across varying state-of-the-art methods and compare their performance. The dataset and code will be available upon publication at \href{https://github.com/NUS-Tim/UNGT}{https://github.com/NUS-Tim/UNGT}.
\end{abstract}


\begin{keyword}
Ultrasound Nasogastric Tube \sep Semi-Supervised Learning \sep Medical Image Analysis



\end{keyword}

\end{frontmatter}



\section{Introduction}
\label{1}

The nasogastric tube (NGT) is a type of tube inserted via the nose into the stomach for feeding, and it has been widely leveraged in clinical practice. The accuracy of NGT placement is vital for patient safety because a misplaced NGT would result in food content entering the lungs, resulting in a chest infection. Conventional bedside techniques such as pH testing and auscultation present unique limitations \cite{fernandez2010accuracy}. For instance, the lack of gastric aspirates may limit pH testing, and the transmitted sounds may interfere with the auscultation. Failure of the bedside methods will require confirmation of NGT placement using a chest X-ray \cite{yildirim2018verifying}. However, this necessitates radiation and transport of patients to the radiology department, which takes time and manpower. Ultrasound (US) has been utilized as an alternative modality to confirm the NGT placement, particularly in acute care settings like the emergency department and intensive care unit \cite{kim2012effectiveness}. Though US scanning can provide direct tube visualization fairly quickly at the bedside, its usage presents varying challenges. For instance, the accuracy of the diagnosis highly depends on the experience, the continuous assessment is time-consuming, and different operators may issue different diagnosis results. To this end, deep learning \cite{ronneberger2015u, wang2020covid} has been introduced to assist in the diagnosis. Although deep learning significantly benefits the US diagnosis, the number and coverage of structures in the US datasets for model training can be largely limited. This can be attributed to the considerable complexity of capturing US images compared with natural images \cite{deng2009imagenet}. This complication arises from the requirement for specialized medical equipment and the involvement of medical experts during the image acquisition process. To our knowledge, existing open-source US datasets concentrating on structures such as breast \cite{yap2020breast,al2020dataset}, head \cite{alzubaidi2023large,lu2022jnu}, thyroid \cite{tnscui2020-seg-rank1st}, and gastrointestinal \cite{he2023query2,jha2023gastrovision}, while there is no readily available US NGT dataset.

Deep learning-based medical image analysis encompasses a range of tasks \cite{liu2023recent,chen2024accurate} such as segmentation, classification, detection, reconstruction, and captioning. Among these, segmentation and classification have emerged as the most prevalently implemented techniques \cite{zhu2024xlstm,wu2025trans,chen2024sckansformer,zhu2025bridging} due to their critical role in facilitating accurate diagnosis and treatment planning. Numerous segmentation approaches adopt UNet and its variants \cite{ronneberger2015u,zhou2018unet++,chen2024transunet} as a foundational architecture. The UNet holds a symmetric design, comprising an encoder for downsampling, a decoder for upsampling, and a bottleneck layer that bridges the encoder and decoder. The incorporation of skip connections facilitates the integration of contextual and spatial information. Several approaches leverage the mean teacher and its derivation as base architecture \cite{yu2019uncertainty, zhang2023uncertainty, wang2023mcf}. The mean teacher architecture consists of a teacher and a student subnet, in which the parameter of the teacher network is updated from the student network through the exponential moving average algorithm. Besides, some methods adopt the architecture with a single encoder paired with multiple decoders \cite{wu2022mutual,tang2023multi,liu2024segmenting} and leverage varying loss function designs to adapt to diverse segmentation targets. The latter two categories are particularly prevalent in semi-supervised learning scenarios, which enables models to be trained effectively with limited data. Despite the advancements in reducing data requirements, these approaches are not specifically designed to address data imbalance. While several studies have attempted to mitigate the issue of unbalanced data \cite{pan2023smile,wang2023dhc}, they can overlook the concurrent challenge posed by limited data. Concerning classification, a significant portion of research leverages ResNet and its variants \cite{he2016deep, lu2020classification, sharma2023brain} as the foundation. The ResNet introduces the residual connection and has notably mitigated the difficulties associated with training deep networks, thus enabling the model to capture rich and discriminative features.

To bridge the dataset gap in the context of US NGT research, we introduce a novel US NGT (UNGT) dataset for medical image analysis. The UNGT dataset comprises 493 US images collected from 110 patients with an average image resolution of around 879 $\times$ 583. Structures including the liver, stomach, tube, and pancreas are precisely annotated. To concurrently address data limitation and imbalance, we introduce a semi-supervised adaptive-weighting aggregation medical segmenter (AAMS). The proposed adaptive weighting (ADW) scheme resolves the severe unbalanced challenge by regulating the loss across varying categories during training. The introduced multiscale attention aggregation (MAA) block enhances the feature representation by incorporating local and global contextual information. Accordingly, the AAMS can prioritize infrequent or minor structures and maintain superior representation ability. The flexibility to handle different unbalanced data and superior representation capability ensure the adaptability and variability of the AAMS. We perform extensive segmentation experiments leveraging AAMS and varying state-of-the-art (SOTA) approaches, and the results demonstrate that the proposed AAMS shows apparent performance leadership. Moreover, we perform comprehensive classification experiments on different SOTA methods and compare their performance. To sum up, our main contributions are:
\begin{itemize}
    \item We introduce an ultrasound nasogastric tube dataset comprising 493 images from 110 patients, featuring an average image resolution of around 879 $\times$ 583. Accurate annotations are performed on the liver, stomach, tube, and pancreas.
    \item We construct a semi-supervised adaptive-weighting aggregation medical segmenter to address data limitation and imbalance in parallel.
    \item We conduct comprehensive segmentation experiments, and the proposed model outperforms prevalent state-of-the-art methods. Besides, we perform extensive classification studies on varying approaches and evaluate their performance.
\end{itemize}

The rest of this paper is organized as follows. Section \hyperref[2]{2} "Related Work" demonstrates the recent progress of medical US datasets and medical image analysis, including segmentation and classification. Section \hyperref[3]{3} "Methods" discusses the development of the UNGT dataset and AAMS model in detail. In Section \hyperref[4]{4} "Experiments", the data preprocessing pipeline, experimental configuration, and evaluation metrics are presented. Detailed experimental results and analysis on the UNGT dataset are in Section \hyperref[5]{5} "Results and Analysis". Section \hyperref[6]{6} "Ablation and Generalization" presents the ablation study of AAMS and evaluates its generalization ability on an open-source dataset. We conclude our research and outline future perspectives in Section \hyperref[7]{7} "Conclusion".


\section{Related Work}
\label{2}

\subsection{Medical Ultrasound Dataset}
\label{2.1}

Numerous US datasets have been developed for medical image analysis on varying structures, such as breast, head, thyroid, and gastrointestinal. For instance, Yap et al. \cite{yap2020breast, yap2017automated} developed a BUS dataset collected from the UDIAT Diagnostic Centre of the Parc Taulí Corporation using the Siemens ACUSON Sequoia C512 system. It comprises 163 images with an average resolution of 760 $\times$ 570, where 110 feature benign lesions and 53 feature cancerous masses. Al et al. \cite{al2020dataset} proposed a BUSI dataset collected from 600 female patients aged between 25 and 75 years using the LOGIQ E9 and LOGIQ E9 Agile US systems. The BUSI comprises 780 images with an average resolution of approximately 500 $\times$ 500, in which 437, 210, and 133 are benign, malignant, and normal. Pedraza et al. \cite{tnscui2020-seg-rank1st, pedraza2015open} constructed a DDTI dataset with 637 B-mode thyroid US images curated at the IDIME US Department in Columbia. The DDTI comprises a variety of lesions such as thyroiditis, cystic nodules, adenomas, and thyroid cancers. Lu et al. \cite{lu2022jnu} proposed an intrapartum transperineal JNU-IFM dataset of the Intelligent Fetal Monitoring Lab of Jinan University. It comprises 6224 US images extracted from 78 videos of 51 pregnant women and encompasses four categories depending on symphysis pubis and fetal head availability. Leclerc et al. \cite{leclerc2019deep} developed a CAMUS dataset with 2000 US images from 500 patients. The images were collected at the University Hospital of St Etienne and annotated in end-systole and end-diastole frames on the left ventricle endocardium, myocardium, and left atrium. Additionally, Alzubaidi et al. \cite{alzubaidi2023large} proposed a large-scale dataset for fetal head biometry with 3832 US images. The dataset highlights the brain, cavum septum pellucidum, and lateral ventricles, and the dimension of the images is 959 $\times$ 661. In 2024, Belghith et al. \cite{belghith2024dataset} constructed a plantar flexor muscle dataset comprising 48 images using an Aixplorer scanner in Shear Wave mode. Half of the images were captured in the prone position, while the remaining were in the quiet standing position.

Several medical datasets have focused on stomach-related US imaging. For example, He et al. \cite{he2023query2} developed a GIST514-DB dataset with 251 leiomyoma images and 263 gastrointestinal stromal tumor lesion images collected in the endoscopy center of the General Hospital of Tianjin Medical University. Minoda et al. \cite{minoda2020efficacy} collected 3980 endoscopic US images from 273 patients with conventional echoendoscopes or mini-probes. In 2023, Jha et al. \cite{jha2023gastrovision} proposed a multi-center gastrointestinal endoscopy dataset termed GastroVision. The GastroVision includes 27 anatomical landmarks and 8000 images collected from Bærum Hospital in Norway and Karolinska University Hospital in Sweden. In addition, Hirai et al. \cite{hirai2022artificial} collected 16110 upper gastrointestinal images from 631 cases at varying hospitals such as the Nagoya University Hospital. The subepithelial lesions include gastrointestinal stromal tumor, leiomyoma, schwannoma, neuroendocrine tumor, and ectopic pancreas. The frequency was set to 12 or 20 MHz, or 5 to 7.5 MHz when the entire image could not be captured. Despite the substantial efforts by researchers to create stomach-related US datasets, there remains a discernible void in datasets tailored for NGT investigations. To address the gap in research endeavors, we introduce a novel UNGT dataset.

\subsection{Medical Image Analysis}
\label{2.2}

Segmentation is one of the most significant tasks in medical image analysis. A plethora of approaches adopt UNet and its variants as the foundation. For instance, Zhou et al. \cite{zhou2018unet++} developed a UNet++ approach, in which the encoder and decoder are connected through nested dense skip pathways. Subsequent work is seen in the UNet 3+ \cite{huang2020unet} with full-scale skip connections and deep supervision. Cao et al. \cite{cao2022swin} constructed a Swin-Unet model, where the tokenized image patches are fed into the transformer-based encoder-decoder to capture local and global semantic features. Chen et al. \cite{chen2024transunet} developed a TransUNet that leverages ViT \cite{dosovitskiy2020image} to encode tokenized patches. The decoder upsamples the encoded features, followed by the integration with convolutional feature maps. Valanarasu \cite{valanarasu2022unext} et al. proposed a UNeXt with an early convolutional stage and a multilayer perceptron stage, in which the tokenized multilayer perceptron block tokenizes and projects the convolutional features. Tagnamas et al. \cite{tagnamas2025sca} developed a SCA-InceptionUNeXt that integrates a modified InceptionNeXt block and a spatial-aware channel attention module into the U-shaped architecture. Ma et al. \cite{ma2024semi} introduced a Semi-Mamba-UNet to integrate a Mamba-based UNet architecture with a conventional convolution-based UNet into the semi-supervised learning framework. Both networks generate pseudo-labels and perform mutual pixel-level supervision. Dong et al. \cite{dong2025uncertainty} proposed an uncertainty-aware consistency learning method to improve generalization and mitigate suboptimal performance in semi-supervised segmentation.

Several methods leverage the mean teacher and its derivation as the base. Yu et al. \cite{yu2019uncertainty} introduced a UA-MT approach where the teacher model generates target outputs along with the uncertainty of each prediction using Monte Carlo sampling. Wang et al. \cite{wang2023mcf} constructed an MCF approach and leveraged a contrastive difference review module to alternative the potential moving average. A rectification loss grounded in the ratio of the potential mispredicted area loss to the area size was also proposed. Wang et al. \cite{wang2022cnn} introduced a S4CVNet, which comprises a feature-learning module and a guidance module. The feature-learning module employs a dual-view co-training strategy that leverages the advantages of the convolutional network and ViT while mitigating architectural discrepancies. Besides, some approaches adopt the architecture with a single encoder paired with multiple decoders and leverage varied loss functions to adapt to diverse segmentation targets. For example, Wu et al. \cite{wu2022mutual} proposed an MC-Net+ with a shared encoder and multiple varying decoders. The proposed method leverages the statistical discrepancy of the decoder outputs to denote model uncertainty. Besides, it introduces a mutual consistency constraint across the probability output and soft pseudo labels across varying decoders. In 2024, we \cite{liu2024segmenting} introduced a DEMS model that features an encoder-decoder architecture and incorporates the online automatic augmenter (OAA) and residual robustness enhancement (RRE) blocks. These approaches leverage the semi-supervised learning scheme, enabling models to be trained effectively with limited data. Despite the advancements in reducing data requirements, these approaches are not tailored to address data imbalance. Several studies have attempted to mitigate the issue of unbalanced data. For example, Pan \cite{pan2023smile} et al. proposed a cost-sensitive learning strategy to solve the imbalanced data distribution by increasing the penalization for the error classification of the minority classes. Wang \cite{wang2023dhc} et al. developed a distribution-aware debiased weighting scheme and a difficulty-aware debiased weighting scheme to guide the model to solve biases. However, these approaches can overlook the concurrent challenge posed by limited data. To jointly handle data limitation and imbalance, we propose a novel AAMS model.

Aside from segmentation, medical image classification serves as an important application scenario. For example, Wang et al. \cite{wang2020covid} introduced a COVID-Net with high architectural diversity and selective long-range connectivity. The intensive utilization of the projection-expansion-projection pattern ensures boosted representational capacity while maintaining computing efficiency. Lu et al. \cite{lu2020classification} proposed a revised ResNet based on the pyramid dilated convolution for Gliomas classification. Specifically, the pyramid dilated convolution was integrated into the bottom of ResNet to achieve a higher receptive field. Sharma et al. \cite{sharma2023brain} constructed an enhanced ResNet by removing the final layer and adding four additional layers for brain tumor detection. Aladhadh et al. \cite{aladhadh2022effective} leveraged the ViT to recognize skin images for skin cancer diagnosis. Images were processed with data augmentation techniques like rotation, flip, contrast, and scaling. In addition, Liang \cite{liang2023light} et al. developed a hybrid lightweight model architecture by connecting convolution layers and MobileViT blocks \cite{mehta2021mobilevit} in tandem. In 2024, Song et al. \cite{song2024cell} introduced a novel worse-case boosting learning algorithm to classify cervical cell images in under-representative cases. The worse-case data was sampled per gradient norm information, and the enhanced loss values were then leveraged to update the classifier. Jim\'enez et al. \cite{jimenez2024gan} leveraged various generative adversarial networks to synthesize US and mammography mass images, enhancing the classification performance of ResNet. Huang et al. \cite{huang2024srt} developed an SRT approach that integrates residual blocks and triplet loss into the SwT \cite{liu2021swin}. This integration enhances the sensitivity to thyroid nodule features and improves the ability to distinguish subtle feature differences.


\section{Methods}
\label{3}

\subsection{Data Acquisition and Characteristics}
\label{3.1}

\begin{figure*}
	\centering
	  \includegraphics[width=\textwidth]{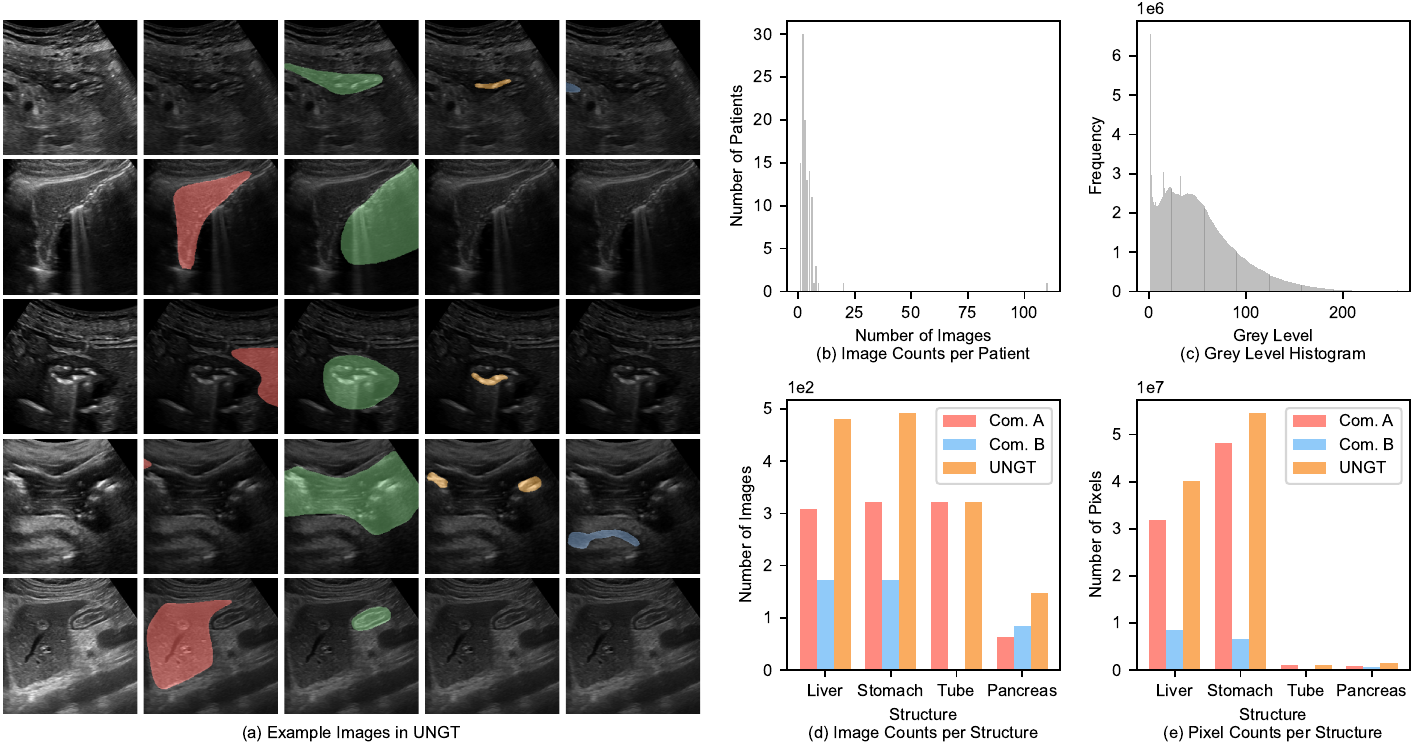}
	\caption{Example images and characteristics analysis of the developed UNGT dataset. Red, green, orange, and blue regions represent the liver, stomach, tube, and pancreas areas. Com. A and Com. B represents the two components. (a) Example images in UNGT, (b) image counts per patient, (c) grey level histogram without zero level, (d) image counts per structure, and (e) pixel counts per structure.}
	\label{fig1}
\end{figure*}

To mitigate the gap from the dataset aspect, we propose a UNGT dataset with 493 US images gathered from 110 patients. We demonstrate several example images within the developed UNGT dataset in \cref{fig1}a. For neatness considerations, images have been resized and cropped as discussed in Section \hyperref[4.1]{4.1}. Of these, 305 images are captured in the transverse section, and 188 are scanned in the longitudinal section. The average resolution of the scans is approximately 879 $\times$ 583. The UNGT dataset comprises precise annotations for the liver, stomach, tube, and pancreas, and we partition it into two components. Specifically, component A encompasses 321 images with the tube collected from 60 patients, and component B consists of 172 images without the tube gathered from 50 patients.

We construct the dataset in two ways. First, we retrospectively trawl the institutional PACS acquired between 1 January 2020 and 31 December 2022 to retrieve pre-existing US images demonstrating the stomach with or without a feeding tube. Second, we prospectively recruit patients with a pre-existing feeding tube in situ for US scanning after the approval of the ethics review board and the acquisition of written patient consent. The IRB number for our work is 2022/00171. We perform the scans using standard commercial US machines, including General Electric LOGIQ E9 and E10, Siemens ACUSON, and Toshiba APLIO 100 on transverse and longitudinal sections. We adjust the curvilinear probe frequency between 2–8 MHz based on the patient's habitus, gain, depth, and focal point on the image for optimal image contrast and resolution. Incorporating images from different US machines introduces heterogeneity into the dataset, enhancing its robustness and real-world applicability. We obtain approximately 2 to 10 static images per patient and capture more frames when the cine-clips are applicable. We review the quality of the US images to ensure adequate clarity, resolution, and field-of-view before selection. A board-certified radiologist with 10 years of experience in US imaging selects the scans, performs cropping to remove confidential information, and carries out the annotation using an open-source tool Labelme \cite{russell2008labelme}. The liver, stomach, tube, and pancreas are annotated, with the tube consistently inside the stomach. Another board-certified radiologist with 5 years of experience in US imaging cross-checks the annotations for accuracy. Annotation discrepancies are resolved by consensus between the two radiologists.

We present the characteristics analysis of our developed UNGT dataset in \cref{fig1}b to \cref{fig1}e. The distribution of image counts per patient reveals that only a limited number of images are typically captured per patient, highlighting the challenges of US image acquisition. The brightness histogram distribution aligns with the typical characteristics of US images, which often exhibit predominantly low average brightness. We exclude the grey level of zero as it can comprise non-scanned regions, such as black triangle borders outside the scan boundary. The primary challenge within the UNGT dataset is the severely unbalanced distribution. Considering the image counts per structure, most images comprise the liver and stomach, while only a small portion involves the tube and pancreas. Such an unbalanced constraint becomes more severe for the size across varying structures. The number of pixels labeled as liver and stomach is not at the same magnitude as that of the tube and pancreas. This can be attributed to the limited number of structures and the small size of such objects. The severely unbalanced distribution at coarse and fine levels demonstrates the difficulty in analyzing the UNGT dataset, especially for the tube and pancreas structures.

\subsection{Model Architecture Design}
\label{3.2}

\begin{figure*}
	\centering
	  \includegraphics[width=\textwidth]{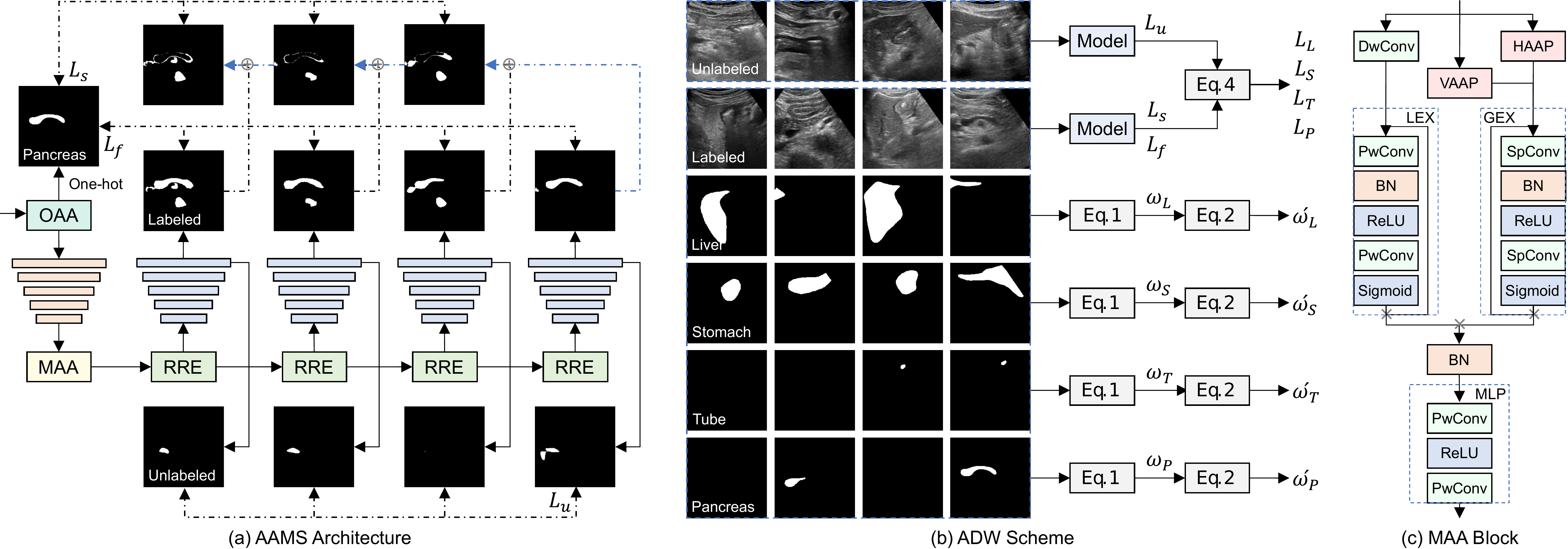}
	\caption{Architecture of the developed AAMS. It comprises an ADW scheme, an MAA block, the OAA, and the RRE blocks. The loss function of AAMS includes fusion loss $L_{f}$, sensitivity loss $L_{s}$, and unsupervised loss $L_{u}$. The ADW scheme calculates each class's normalized loss weight $\omega'$. The MAA block is mainly composed of a local excitation (LEX) block, a global excitation (GEX) block, and a multi-layer perceptron (MLP) block. $\oplus$ and $\times$ denote the exclusive OR and multiplication operations. L, S, T, and P stand for the liver, stomach, tube, and pancreas. HAAP and VAAP represent horizontal and vertical adaptive average pooling layers. DwConv, SpConv, and PwConv present depthwise, stripe, and pointwise convolution layers. BN stands for batch normalization layer. (a) AAMS architecture, (b) ADW scheme, and (c) MAA block.}
	\label{fig2}
\end{figure*}

To tackle data limitation and imbalance jointly, we propose an AAMS based on our DEMS. The DEMS features an encoder-decoder architecture and incorporates the OAA and RRE blocks. The OAA augments the input with various image transformations to diversify the dataset for generalization ability. The RRE elevates feature diversity and introduces perturbations to create different inputs for varying decoders for enhanced variability. Moreover, DEMS incorporates a novel sensitivity loss $L_{s}$ to enhance decoder consistency and stabilize training. The formulation of the sensitivity loss is based on an intuitive concept in binary segmentation where a mismatch in predictions across two decoders suggests an error in one.

We advance the DEMS to the multi-category segmentation context using parallel binary segmentation and propose an ADW scheme and an MAA block. We present the detailed architecture of the proposed AAMS in \cref{fig2}. It comprises an ADW scheme, an MAA block, the OAA, and the RRE blocks. The loss function of AAMS comprises fusion loss $L_{f}$, $L_{s}$, and unsupervised loss $L_{u}$. The $L_{f}$ is composed of BCE loss and dice loss, while the $L_{u}$ is based on MSE loss. The ADW scheme tackles class imbalance by adaptively regulating the loss contributions across various categories during training. The MAA block enhances feature representation by integrating local and global contextual representations.

The ADW scheme computes the number of pixels for each category for each batch. Given a batch of encoded mask $M$, the labeled batch size $B$, the number of categories $C$, and the visual dimension $H \times W$, the loss weight $\omega$ for each class can be formulated though \cref{eq1}:
\begin{equation}
\omega_{c} = 1 - \sqrt{\frac{\sum_{b=1}^{B} \sum_{h=1}^{H} \sum_{w=1}^{W} I({M}_{b, c, h, w}=1)}
{B \times H \times W}}\,,
\label{eq1}
\end{equation}

where $I(\cdot)$ represents the indicator function, which returns 1 if the condition is true and 0 otherwise. Specifically, $I({M}_{b, c, h, w}=1)$ indicates whether a pixel located at position $(h, w)$ in the $b$-th image of a batch belongs to category $c$. The calculated category weights are then normalized to ensure the sum of the weights equals 1 using \cref{eq2}:
\begin{equation}
\omega_{c}' = \frac{\omega_c}{\sum_{c=1}^{C} \omega_c}\,.
\label{eq2} 
\end{equation}

The Gaussian warming-up function $\lambda$ is adapted to ensure enhanced training stability. It is formulated based on the training step $t$ and maximum training step $t_{max}$ and can be expressed in \cref{eq3}:
\begin{equation}
\lambda(t, t_{max}) = 
\begin{cases} 
1 & t_{max} = 0\,, \\
\exp\left(-5 \left(1 - \frac{t}{t_{max}}\right)^2\right) & 0 \leq t \leq t_{max}\,.
\end{cases}
\label{eq3}
\end{equation}

Each category loss $L_{c}$ comprises $L_{f}$ and $L_{s}$ for labeled images, $L_{u}$ for unlabeled images, and the Gaussian warming-up function, as depicted in \cref{eq4}:
\begin{equation}
L_c = {L}_{f} + \lambda ({L}_{s} + {L}_{u})\,.
\label{eq4}
\end{equation}

The loss function of AAMS $L$ can be formulated through the multiplication and summation of each category loss and their corresponding normalized weights, as described in \cref{eq5}:
\begin{equation}
L = \sum_{c=1}^{C} \omega_{c}' L_c\,.
\label{eq5} 
\end{equation}

The MAA block primarily comprises a local excitation (LEX) block, a global excitation (GEX) block, and a multi-layer perceptron (MLP) block. From a scale perspective, it comprises a local and a global branch to extract features across multiple scales. In the local branch, a depthwise convolution layer captures local features and transmits them to the LEX block to refine the attention to significant regions. The LEX block integrates two pointwise convolution (PwConv) layers. The first is succeeded by a batch normalization layer and ReLU activation, and the second by a Sigmoid activation. The resulting attention is multiplicatively applied to the input features to scale them by the learned importance. For the global branch, the horizontal adaptive average pooling (HAAP) and vertical adaptive average pooling (VAAP) layers aggregate the context in horizontal and vertical directions. The aggregated features from the HAAP and VAAP layers are fed to the GEX block. The GEX block is structurally similar to the LEX block, but the PwConv layers are replaced with stripe convolution layers in orthogonal directions to better adapt to global attention. The processed local and global contextual features are combined multiplicatively, followed by a batch normalization layer and an MLP block for further representation enhancement. The MLP block consists of two PwConv layers with a ReLU activation layer in between.


\section{Experiments}
\label{4}

\subsection{Dataset Preprocessing}
\label{4.1}

We resize and crop the images and masks to 224 $\times$ 224 resolution. As the tube location varies and holds a small size, directly leveraging the center crop may lead to the entire tube being cropped out. To this end, we propose a dynamic crop approach based on the tube location. Firstly, we resize the images and mask and keep the aspect ratio unchanged to ensure the minimum side equals 224. Secondly, we detect the tube location and set the average tube position to the crop center. The center of the image and mask is regarded as the crop center in case the tube is not detected. Finally, we adjust the location of the crop area to ensure that the cropped area is not out of bounds and perform the crop.

\subsection{Configuration and Evaluation}
\label{4.2}

We conduct extensive segmentation and classification experiments to provide comprehensive and convincing evaluations. For segmentation, we segment the four structures in the UNGT dataset, including the liver, stomach, tube, and pancreas. Regarding classification, we classify the existence of the tube within each image. We leverage PyTorch as the training framework. We report the mean and standard deviation of three independent runs with varying seeds in percentage and present the standard deviation in superscript.

Regarding segmentation, we one-hot encode the mask, ensuring the tube location is encoded to both the stomach and tube channel. We partite the dataset into a training and a validation subset with a ratio of 7:3 and leverage 80\% labeled and 20\% unlabeled images to train semi-supervised approaches. We set SGD as the optimizer and assign the momentum and weight decay to 0.9 and 0.0001. We assign the initial learning rate to 0.01. We leverage the cosine annealing schedule \cite{loshchilov2016sgdr} to update the learning rate during training AAMS. We set the maximum training iterations to 20000 and the batch size to 8, with an identical labeled and unlabeled batch size of 4 for semi-supervised approaches. The loss functions and data augmentation strategies vary across methods. We evaluate the model performance leveraging dice score (DSC), sensitivity (SEN), and precision (PRE).

Considering classification, we divide the dataset into a training subset and a validation subset with a ratio of 7:3. We maintain the ratio of examples with and without the tube in the training and validation subsets, ensuring the class balance across each category. We leverage Adam as the optimizer and set the weight decay to 0.001. We assign the initial learning rate to 0.01 and decrease it by a factor of 0.5 and patience of 5 as training proceeds. We set the maximum epoch to 200 and the batch size to 16. We leverage the cross-entropy loss as the loss function and perform data augmentation through the OAA. Metrics for performance evaluation include accuracy (ACC), SEN, PRE, specificity (SPE), negative predictive value (NPV), and F-1 score (FOS).


\section{Results and Analysis}
\label{5}

\subsection{Segmentation}
\label{5.1}

\begin{table*}[ht]
\centering
\caption{Segmentation results across AAMS and SOTA approaches on the developed UNGT dataset. L, S, T, and P present the liver, stomach, tube, and pancreas. $^{\dagger}$ denotes semi-supervised approach.}
\resizebox{\linewidth}{!}{  
\begin{tabular*}{736pt}{c|cccc|cccc|cccc}  
\toprule
    \multirow{2}{*}{Model}                        & \multicolumn{4}{c|}{DSC}                                                              & \multicolumn{4}{c|}{SEN}                                                              & \multicolumn{4}{c}{PRE}                                                               \\
                                                  & L & S & T & P                                                                         & L & S & T & P                                                                         & L & S & T & P                                                                         \\
\midrule
    UNet \cite{ronneberger2015u}                  & 67.07\ts{0.73}      & 73.12\ts{0.14}      & 29.92\ts{0.07}      & 18.13\ts{0.53}      & 67.84\ts{0.87}      & 74.85\ts{1.60}      & 30.15\ts{0.39}      & 17.93\ts{0.75}      & 71.79\ts{1.04}      & 78.58\ts{1.64}      & 35.36\ts{0.82}      & 20.79\ts{0.67}      \\
    TransUnet \cite{chen2024transunet}            & 68.20\ts{0.07}      & 75.13\ts{0.13}      & 33.43\ts{1.21}      & 17.75\ts{0.04}      & 69.50\ts{0.82}      & 79.04\ts{0.39}      & 37.86\ts{2.95}      & 17.40\ts{0.81}      & 72.02\ts{0.56}      & 77.63\ts{0.50}      & 35.70\ts{1.79}      & 20.65\ts{0.26}      \\
    UNext \cite{valanarasu2022unext}              & 65.47\ts{0.64}      & 65.65\ts{0.94}      & 20.45\ts{0.44}      & 13.82\ts{0.74}      & 65.90\ts{0.58}      & 66.96\ts{0.89}      & 19.58\ts{0.55}      & 12.52\ts{0.93}      & 70.22\ts{1.70}      & 72.29\ts{1.43}      & 27.27\ts{1.05}      & 18.70\ts{1.06}      \\
    CMUNeXt \cite{tang2024cmunext}                & 67.73\ts{0.95}      & 73.52\ts{0.31}      & 32.86\ts{1.38}      & 17.78\ts{0.82}      & 67.88\ts{1.65}      & 74.64\ts{0.87}      & 35.78\ts{1.99}      & 17.44\ts{0.76}      & 73.02\ts{0.80}      & 79.15\ts{1.14}      & 35.62\ts{0.77}      & 20.95\ts{0.21}      \\
    MedT \cite{valanarasu2021medical}             & 59.85\ts{1.75}      & 63.91\ts{1.12}      & 4.16\ts{2.50}       & 9.44\ts{0.78}       & 57.83\ts{2.98}      & 66.57\ts{1.71}      & 2.87\ts{1.75}       & 8.38\ts{0.98}       & 67.90\ts{0.28}      & 69.46\ts{0.45}      & 12.54\ts{6.02}      & 14.75\ts{1.11}      \\
    CMUNet \cite{tang2023cmu}                     & 71.04\ts{0.20}      & 76.42\ts{0.16}      & 35.84\ts{1.12}      & 19.64\ts{0.10}      & 71.35\ts{0.68}      & 78.48\ts{1.13}      & 37.30\ts{1.41}      & 19.20\ts{0.22}      & 74.48\ts{1.65}      & 80.41\ts{0.82}      & 39.59\ts{0.76}      & 21.34\ts{0.56}      \\
    CCT$^{\dagger}$ \cite{ouali2020semi}          & 67.51\ts{0.94}      & 72.68\ts{0.91}      & 33.96\ts{1.19}      & 18.86\ts{0.06}      & 67.43\ts{1.76}      & 76.69\ts{0.76}      & 37.11\ts{3.49}      & \tb{20.04}\ts{0.42} & 73.83\ts{0.97}      & 76.63\ts{1.01}      & 36.67\ts{1.15}      & 19.15\ts{0.77}      \\
    CPS$^{\dagger}$ \cite{chen2021semi}           & 69.22\ts{1.50}      & 72.70\ts{0.27}      & 35.78\ts{0.26}      & 19.25\ts{0.61}      & 70.19\ts{2.69}      & 75.79\ts{0.67}      & 36.31\ts{0.39}      & 19.68\ts{1.17}      & 73.09\ts{0.47}      & 76.71\ts{1.42}      & 39.48\ts{0.36}      & 20.90\ts{0.36}      \\
    CTCT$^{\dagger}$ \cite{luo2022semi}           & 72.29\ts{0.40}      & 75.20\ts{0.40}      & 38.28\ts{0.69}      & 19.77\ts{0.06}      & \tb{73.79}\ts{0.82} & 77.02\ts{0.80}      & 39.23\ts{1.28}      & 19.73\ts{0.25}      & 74.84\ts{0.93}      & 79.63\ts{0.90}      & 42.00\ts{0.57}      & 21.14\ts{0.24}      \\
    ICT$^{\dagger}$ \cite{verma2022interpolation} & 69.38\ts{1.29}      & 73.24\ts{0.71}      & 34.54\ts{0.73}      & 18.53\ts{0.07}      & 69.82\ts{1.88}      & 76.61\ts{1.40}      & 34.94\ts{1.05}      & 18.50\ts{0.38}      & 74.66\ts{0.16}      & 77.55\ts{0.28}      & 37.94\ts{0.59}      & 20.44\ts{0.33}      \\
    R-Drop$^{\dagger}$ \cite{wu2021r}             & 71.49\ts{0.52}      & 73.23\ts{0.71}      & 33.52\ts{1.05}      & 19.48\ts{0.62}      & 72.81\ts{1.03}      & 76.55\ts{2.33}      & 32.95\ts{1.83}      & 19.58\ts{1.22}      & 74.63\ts{0.28}      & 77.57\ts{1.06}      & 38.51\ts{0.79}      & 21.38\ts{0.48}      \\
    URPC$^{\dagger}$ \cite{luo2021efficient}      & 68.77\ts{1.51}      & 73.19\ts{0.16}      & 32.60\ts{0.45}      & 19.30\ts{0.29}      & 69.48\ts{2.52}      & 75.35\ts{1.03}      & 33.10\ts{0.37}      & 19.04\ts{0.46}      & 73.43\ts{0.47}      & 77.83\ts{1.00}      & 36.17\ts{0.26}      & 21.07\ts{0.28}      \\
    MGCC$^{\dagger}$ \cite{tang2023multi}         & 70.25\ts{0.40}      & 75.86\ts{0.57}      & 28.23\ts{0.39}      & 17.14\ts{0.60}      & 69.93\ts{0.65}      & 75.39\ts{0.80}      & 29.57\ts{1.23}      & 15.98\ts{0.69}      & 76.18\ts{1.18}      & \tb{82.87}\ts{0.08} & 32.32\ts{0.64}      & 21.89\ts{0.41}      \\
    AAMS$^{\dagger}$ (Ours)                       & \tb{72.74}\ts{0.81} & \tb{78.55}\ts{0.06} & \tb{41.07}\ts{0.74} & \tb{20.74}\ts{0.49} & 73.00\ts{1.16}      & \tb{81.67}\ts{0.58} & \tb{42.63}\ts{0.74} & 19.98\ts{0.44}      & \tb{77.94}\ts{0.80} & 80.94\ts{0.36}      & \tb{43.04}\ts{0.59} & \tb{23.19}\ts{0.21} \\
\bottomrule
\end{tabular*}
}
\label{tab1}
\end{table*}

We present the segmentation results across AAMS and SOTA methods on the developed UNGT dataset in \cref{tab1}. As can be seen, the introduced AAMS represents superior performance leadership compared with SOTA methods to varying extents on four structures. Regarding the liver, the AAMS achieves the highest DSC and PRE of 72.74\% and 77.94\%. Though CTCT presents the optimal SEN of 73.79\%, it does not exhibit an extraordinary PRE of 74.84\%. The MedT does not present exemplary performance under the current configuration with a DSC of 59.85\%. Considering the stomach, the AAMS realizes the optimal DSC and SEN of 78.55\% and 81.67\%. Although MGCC archives the highest PRE, it achieves a relatively low SEN of 75.39\%. Analogously, the MedT demonstrates an unexpected performance with a DSC equal to 63.91\%. Significant performance decreases are identified in the results of the tube segmentation, preeminently attributed to the unbalanced constraint within the dataset. Despite this, the AAMS realizes lower performance degradation across varying approaches and achieves the optimal DSC, SEN, and PRE of 41.07\%, 42.63\%, and 43.04\%. Greater performance degradation is observed for most SOTA methods, and MedT realizes an extremely low DSC of merely 4.16\%. Analogously, further performance decreases are observed in pancreas segmentation due to the severe limitation on the number of images. Despite this, the AAMS realizes the optimal DSC and PRE of 20.74\% and 23.19\%. Though CCT achieves the optimal SEN of 20.04\%, AAMS follows closely with a performance of 19.98\%. The lowest DSC of 9.44\% is achieved by MedT, demonstrating higher performance compared with the tube. This suggests that MedT may not be well-suited for segmenting extremely minor objects. The performance difference among varying structures reflects the difficulty of segmenting the unbalanced UNGT dataset, and the metrics leadership of the proposed AAMS emphasizes its adaptability and variability.

\begin{figure*}
	\centering
	  \includegraphics[width=\textwidth]{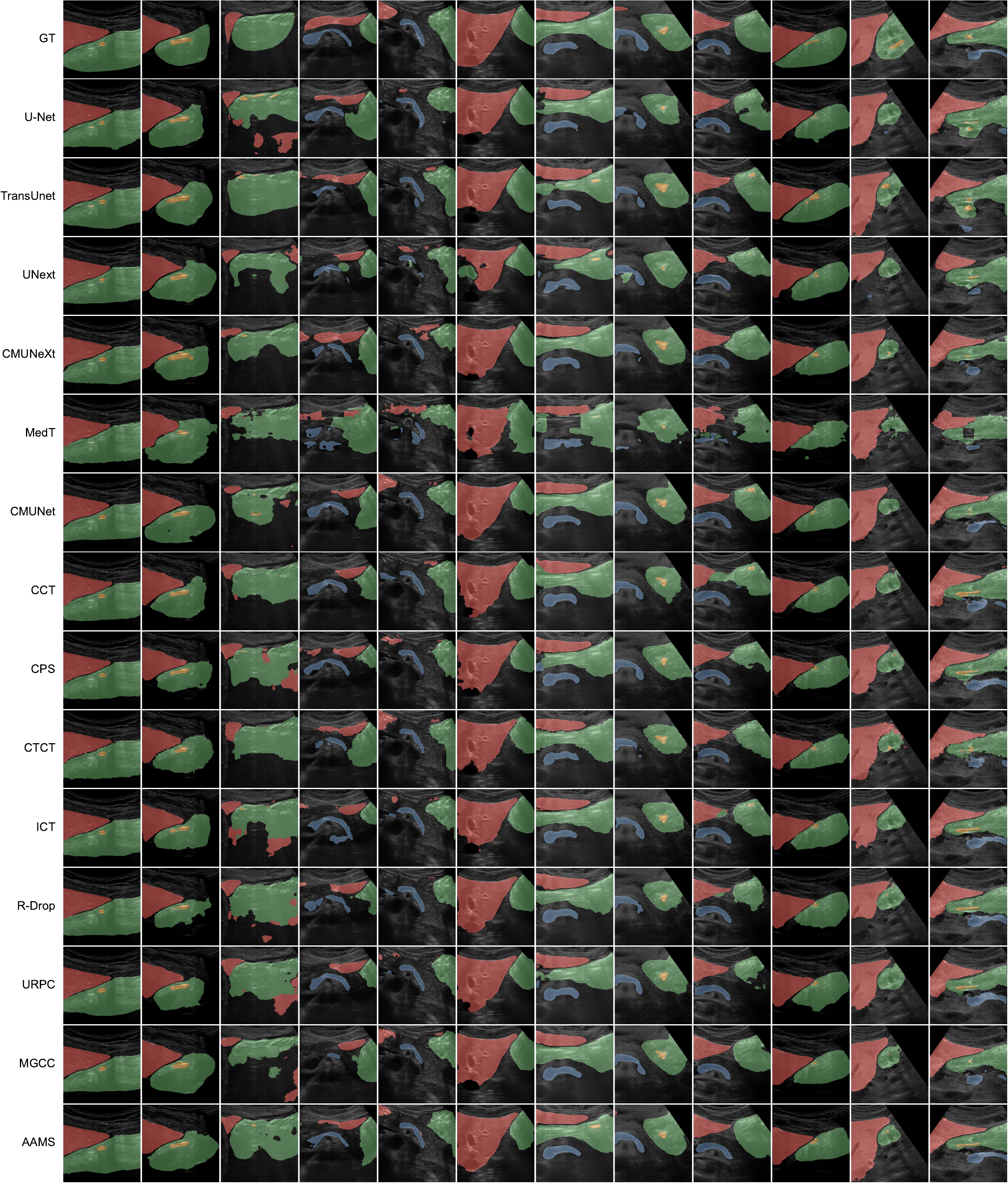}
	\caption{Predicted and GT masks across AAMS and SOTA approaches on the introduced UNGT dataset. GT stands for ground truth.}
	\label{fig3}
\end{figure*}

We demonstrate the predicted and ground truth (GT) masks across AAMS and SOTA approaches on the introduced UNGT dataset in \cref{fig3}. Results indicate that the AAMS demonstrates boosted prediction accuracy compared with varying SOTA approaches. In the first image, the AAMS effectively prevents the merging of adjacent structural areas and ensures accurate delineation, while approaches like ICT output the liver and stomach areas without a clear boundary. For the fifth image, the AAMS predicts the most comprehensive pancreas areas, while approaches such as CCT and R-Drop present a fragmented pancreas. Concerning the seventh image, the AAMS outputs the most accurate stomach shape despite the stomach shape deviating from common forms. Meanwhile, models such as UNext and URPC predict undesirable stomach shapes. As for the eighth figure, the AAMS predicts the tube and pancreas areas more accurately, whereas approaches such as UNext yield an imprecise tube, and CMUNeXt outputs a fragmented pancreas. Analogously, the MedT demonstrates the most deficient prediction areas. A marked proportion of its prediction regions are fragmented, featuring unreasonable structure boundaries identified. For instance, rectangle boundaries are observed within the stomach structure in the seventh and last images. It is worth noting that most of the approaches tend to wrongly predict the light region as a tube due to their analogous characteristics. An intuitive example is shown in the last image, where the light-long strip region is erroneously identified as a tube by most methods. The vast number of wrongly predicted regions emphasizes the difficulty of segmenting the introduced UNGT dataset, and the relatively superior prediction accuracy demonstrates the effectiveness of the developed AAMS.

\begin{figure*}
	\centering
	  \includegraphics[width=\textwidth]{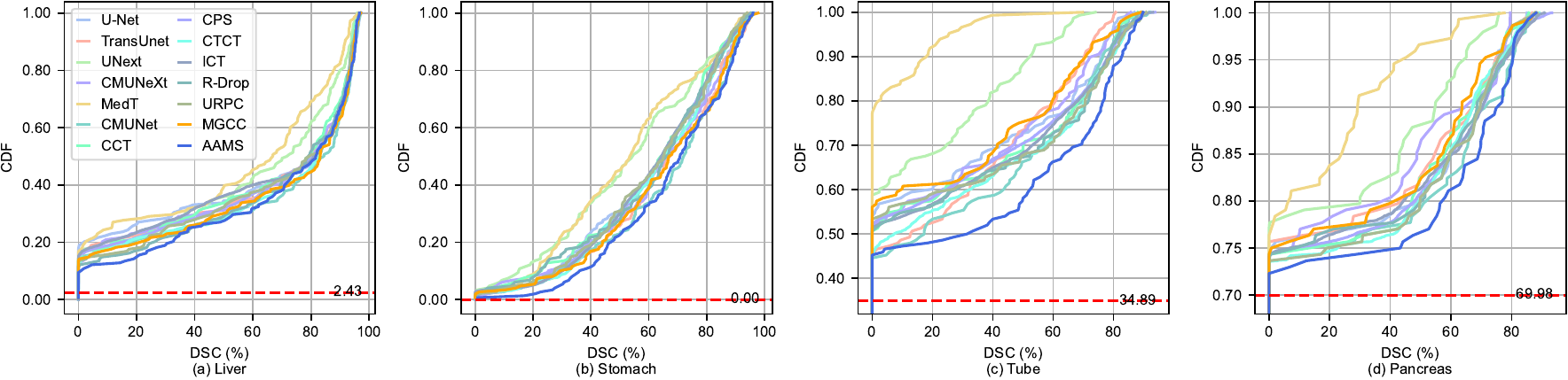}
	  \caption{Cumulative distribution function of varying structures regarding DSC across different methods on the developed UNGT dataset. The horizontal red dotted lines depict the proportion of the GT mask without the corresponding structures in percentage. (a) Liver, (b) stomach, (c) tube, and (d) pancreas.}
	  \label{fig4}
\end{figure*}

We visualize the cumulative distribution function across different structures regarding DSC for varying methods on the developed UNGT dataset in \cref{fig4}. The red dotted lines represent the proportion of the GT mask in which the corresponding structure does not exist in percentage. The predictions below the red dotted lines may not be explained as inaccurate as the DSC can equal zero in case both GT and prediction masks do not present the structure. Through identification, it can be found that the AAMS demonstrates superior overall performance across the four structures. Considering the liver, AAMS presents as one of the optimal approaches, while MedT and UNext reveal relatively undesirable results. Similar observations can be seen in the results of stomach segmentation, where most methods demonstrate desirable performance except for MedT and UNext. Concerning the tube, the AAMS demonstrates tremendous performance leadership compared with the remaining approaches. Under this challenging scenario, the performance of MedT and UNext deteriorates dramatically, and most of the tube prediction of MedT presents a DSC of zero. The trends of the pancreas conform to that of the tube, in which the AAMS demonstrates apparent leadership while MedT and UNext fall behind. The undesirable results presented in the tube and pancreas segmentation reveal the difficulty in segmenting the unbalanced UNGT dataset, and the substantial performance leadership on these structures illustrates the priority of the introduced AAMS.

\begin{figure*}
	\centering
	  \includegraphics[width=\textwidth]{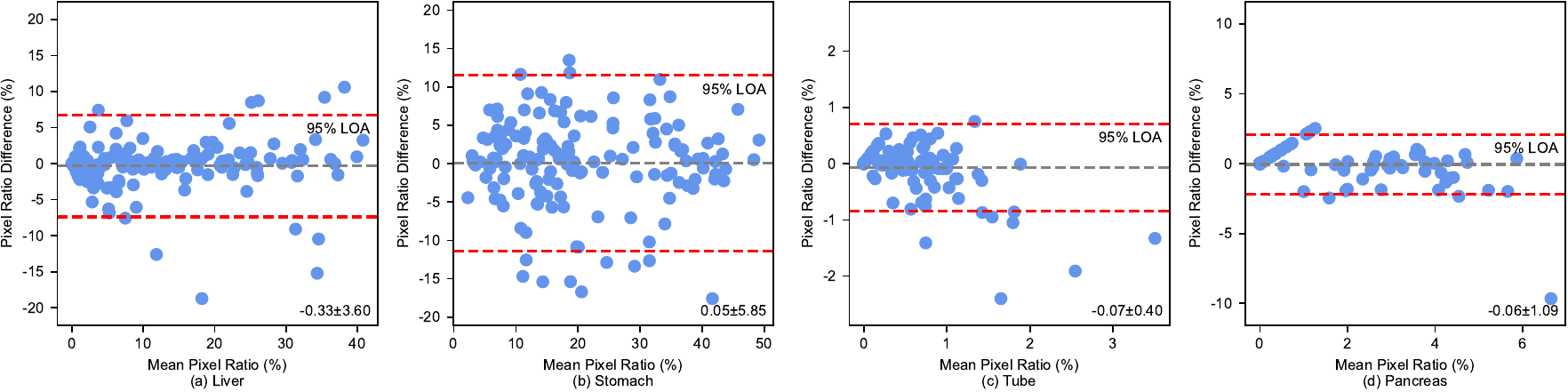}
	\caption{Bland-Altman analysis of AAMS across varying structures on the developed UNGT dataset. LOA demonstrates the limits of agreement. (a) Liver, (b) stomach, (c) tube, and (d) pancreas.}
	\label{fig5}
\end{figure*}

We present the Bland-Altman plot of AAMS across varying structures on the introduced UNGT dataset in \cref{fig5}. We calculate the ratio of each structure pixel in the predicted and GT masks by counting the number of pixels and dividing it by $224^2$. The consistency across each prediction-GT pair is then compared. From the results, the prediction of AAMS depicts superior consistency compared with the GT. The mean differences are overall close to 0\%, with the maximal and minimal differences equal to 0.33\% and 0.05\% for liver and stomach. The standard deviations lie within the expected range and vary with the size of different structures. Explicitly, the highest and lowest standard deviations of 5.85\% and 0.40\% are identified for the stomach and tube. Besides, a dominant portion of the points is located within the 95\% limits of agreement. Though several outliers are observed for each structure, the number of outliers aligns with the expectation and can be attributed to the limited size of the UNGT dataset. In a dataset with a few hundred images, certain patterns may be present in the validation subset but absent or underrepresented in the training subset. Notably, most of the outliers of the tube and pancreas lie along the negative axis, indicating that the AAMS consistently tends to underpredict these structures. This observation conforms to our expectations since most of these structures are relatively small, leading the network to underestimate the larger ones. In this case, the difference across predicted and GT masks is primarily influenced by the structure-specific attributes instead of the inherent constraints of the AAMS. Besides, plots across varying structures evince no uniform difference trend, demonstrating that the observed discrepancies lack a stringent systematic pattern. The high consistency among the prediction-GT mask pairs reveals the effectiveness of the proposed AAMS.

\subsection{Classification}
\label{5.2}

\begin{table}[ht]
\centering
\caption{Classification results across varying SOTA approaches on the introduced UNGT dataset.}
\resizebox{0.6\linewidth}{!}{  
\begin{tabular*}{428pt}{c|ccccccc}  
\toprule
    Model                                    & ACC                 & SEN                 & PRE                 & SPE                 & NPV                 & FOS                 \\
\midrule
    ResNet \cite{he2016deep}                 & 84.10\ts{1.27}      & 86.60\ts{0.82}      & 88.73\ts{1.22}      & 79.47\ts{2.41}      & 76.07\ts{1.56}      & 87.63\ts{0.97}      \\
    InceptionNext \cite{yu2024inceptionnext} & 72.27\ts{1.36}      & 70.47\ts{4.31}      & 84.47\ts{1.62}      & 75.60\ts{4.53}      & 58.03\ts{2.30}      & 76.70\ts{1.92}      \\
    RegNetX \cite{radosavovic2020designing}  & 80.53\ts{0.53}      & 86.27\ts{2.57}      & 84.33\ts{2.08}      & 69.87\ts{5.49}      & 73.37\ts{2.37}      & 85.23\ts{0.41}      \\
    GhostNet \cite{han2020ghostnet}          & 85.67\ts{2.23}      & 88.63\ts{2.92}      & \tb{89.30}\ts{1.88} & \tb{80.13}\ts{3.95} & 79.30\ts{4.03}      & 88.97\ts{1.80}      \\
    CECT \cite{liu2024cect}                  & \tb{85.90}\ts{3.43} & \tb{91.07}\ts{4.81} & 87.70\ts{1.00}      & 76.27\ts{0.90}      & \tb{82.73}\ts{7.82} & \tb{89.33}\ts{2.86} \\
    CSPNet \cite{wang2020cspnet}             & 77.17\ts{2.37}      & 78.33\ts{4.67}      & 85.47\ts{1.80}      & 75.03\ts{4.15}      & 65.40\ts{4.47}      & 81.67\ts{2.36}      \\
    MobileViT \cite{mehta2021mobilevit}      & 82.80\ts{4.08}      & 86.93\ts{3.38}      & 86.63\ts{3.07}      & 75.00\ts{5.67}      & 75.53\ts{6.05}      & 86.77\ts{3.18}      \\
    LeViT \cite{graham2021levit}             & 84.80\ts{2.07}      & 90.07\ts{1.29}      & 87.23\ts{3.42}      & 75.03\ts{8.16}      & 80.17\ts{0.69}      & 88.53\ts{1.25}      \\
    MobileNetV4 \cite{qin2025mobilenetv4}    & 82.37\ts{0.33}      & 84.90\ts{1.75}      & 87.67\ts{1.72}      & 77.57\ts{3.95}      & 73.43\ts{1.24}      & 86.23\ts{0.21}      \\
    RepViT \cite{wang2024repvit}             & 82.77\ts{2.19}      & 89.67\ts{3.65}      & 84.77\ts{1.45}      & 69.87\ts{3.63}      & 78.97\ts{6.21}      & 87.10\ts{1.84}      \\
\bottomrule
\end{tabular*}
}     
\label{tab5}
\end{table}

We appraise the classification performance among different SOTA methods on the proposed UNGT dataset in \cref{tab5}. An insightful examination delineates that the GhostNet and CECT showcase superior performance compared to the remaining SOTA approaches. Specifically, CECT presents the highest ACC, SEN, NPV, and FOS of 85.90\%, 91.07\%, 82.73\%, and 89.33\%, and GhostNet demonstrates the optimal PRE and SPE of 89.30\% and 80.13\%. Conversely, the InceptionNext demonstrates somewhat undesirable results with an ACC of 72.27\%. It is worth noting that the standard deviations observed across varying methods are relatively pronounced, which may be ascribed to the constrained size of the introduced UNGT dataset. Despite this, approaches such as MobileNetV4 and RegNetX demonstrate ideal standard deviations. Taking ACC as an example, these two approaches present standard deviations of 0.33\% and 0.53\%, demonstrating remarkable training stability under the small data regime. In contrast, MobileViT exhibits a significantly elevated ACC standard deviation of 4.08\%, evidencing potential dramatic volatility as training proceeds. The exceptional performance metrics achieved by CECT and GhostNet underscore their effectiveness, while the outstanding standard deviations exhibited by MobileNetV4 and RegNetX emphasize their superior training stability.

\begin{figure*}
	\centering
	  \includegraphics[width=\textwidth]{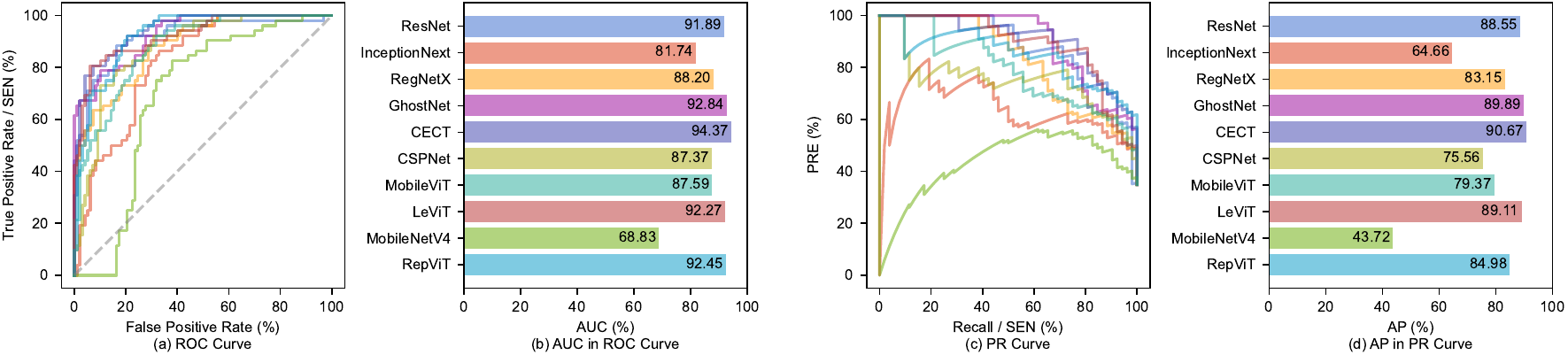}
	  \caption{Receiver operating characteristic (ROC) curve and precision-recall (PR) curve across varying SOTA approaches on the proposed UNGT dataset. The false positive rate presents the proportion of negative samples incorrectly predicted as positive. AUC shows the area under the curve in the ROC curve, and AP represents the average precision within the PR curve. (a) ROC curve, (b) AUC in the ROC curve, (c) PR curve, and (d) AP in the PR curve.}
	  \label{fig6}
\end{figure*}

We present the receiver operating characteristic (ROC) curve and precision-recall (PR) curve among different SOTA approaches on the introduced UNGT dataset in Figure \cref{fig6}. Observations reveal that CECT and GhostNet outperform the remaining approaches. In the ROC curve, CECT and GhostNet achieve an apex area under the curve (AUC) of 94.37\% and 92.84\%. The RepViT and LeViT closely follow those two with a desirable AUC of 92.45\% and 92.27\%. Despite achieving relatively high metrics, MobileNetV4 presents the lowest 68.83\% AUC. High metrics with low AUC suggest that MobileNetV4 struggles with overall discrimination across varying thresholds, potentially limiting its generalization and robustness. Considering the PR curve, CECT and GhostNet consistently realize desirable average precision (AP) of 90.67\% and 89.89\% despite the overall performance decrease compared with the ROC curve. Likewise, MobileNetV4 does not present a preferable AP of 43.72\%. This aligns with its AUC observation, indicating challenges in maintaining a consistent balance across metrics. The comparative subperformance on the PR curve proves the difficulty of classifying US images under a small data regime, and the performance leadership of CECT and GhostNet demonstrates their superiority in handling this scenario.


\section{Ablation and Generalization}
\label{6}

\subsection{Ablation Study}
\label{6.1}

\begin{figure*}
	\centering
	  \includegraphics[width=\textwidth]{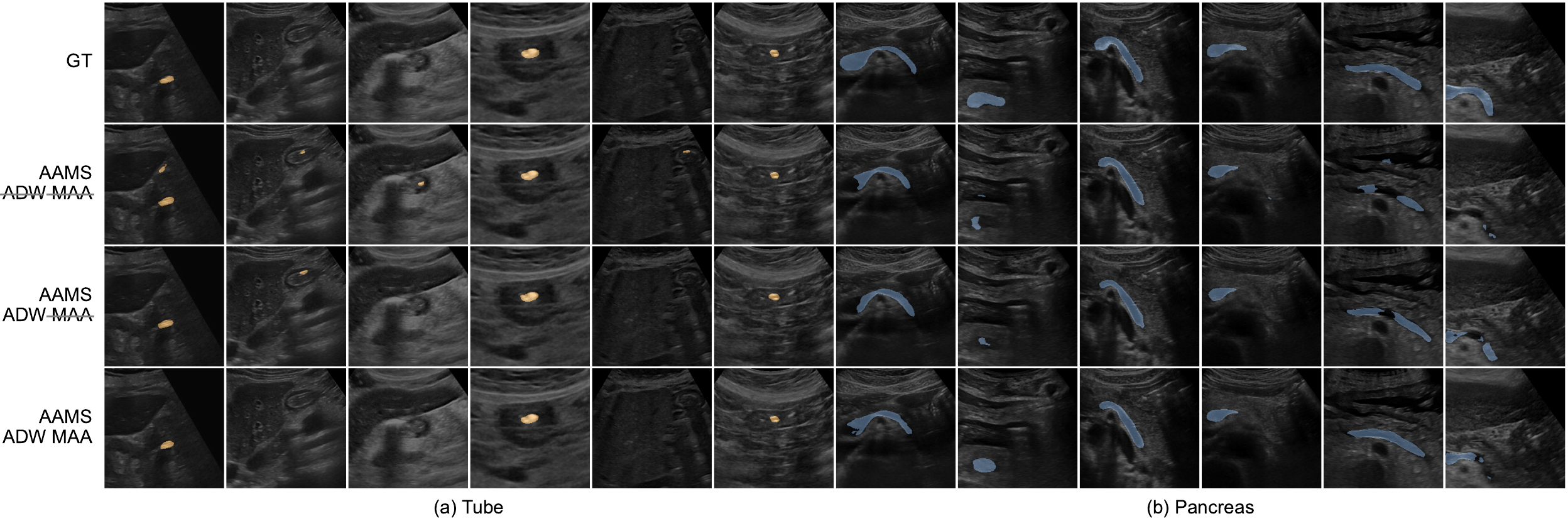}
	\caption{Ablation study of the ADW scheme and MAA block across the tube and pancreas on the introduced UNGT dataset. (a) Tube, and (b) pancreas.}
	\label{fig7}
\end{figure*}

We illustrate the ablation study of the ADW scheme and MAA block across the tube and pancreas on the proposed UNGT dataset in \cref{fig7}. The results indicate that prediction accuracy decreases with the removal of a model component, and removing both components leads to a more severe performance drop. For the second image, the AAMS correctly identifies the absence of the tube structure, while its two variants fail to do so. As for the fifth figure, AAMS consistently captures the tube absence, whereas its variant without the ADW scheme and MAA block fails in this regard. Regarding the eighth image, the AAMS produces the most accurate pancreas shape, while its two ablation variants substantially underestimate the pancreas size. In the eleventh figure, the AAMS delivers the most complete pancreas, whereas both ablation variants output fragmented representations, particularly the variant without both the ADW scheme and MAA block. The performance degradation after removing different numbers of components underscores their importance in the proposed AAMS.

\subsection{Generalization Evaluation}
\label{6.2}

\begin{table*}[ht]
\centering
\caption{Segmentation results across AAMS and SOTA approaches on the CAMUS dataset. LV\textsubscript{endo}, LV\textsubscript{epi}, and LA present the left ventricle endocardium, epicardium contour, and left atrium.}
\resizebox{0.8\linewidth}{!}{  %
\begin{tabular*}{574pt}{c|ccc|ccc|ccc}  
\toprule
    \multirow{2}{*}{Model}                        & \multicolumn{3}{c|}{DSC}                                        & \multicolumn{3}{c|}{SEN}                                        & \multicolumn{3}{c}{PRE}                                         \\
                                                  & LV\textsubscript{endo} & LV\textsubscript{epi} & LA             & LV\textsubscript{endo} & LV\textsubscript{epi} & LA             & LV\textsubscript{endo} & LV\textsubscript{epi} & LA             \\
\midrule
    UNet \cite{ronneberger2015u}                  & 93.73\ts{0.02}      & 87.70\ts{0.09}      & 89.08\ts{0.05}      & 94.08\ts{0.17}      & 87.91\ts{0.29}      & 89.33\ts{0.40}      & 93.82\ts{0.14}      & 87.91\ts{0.12}      & 90.38\ts{0.40}      \\
    TransUnet \cite{chen2024transunet}            & 93.43\ts{0.03}      & 87.55\ts{0.15}      & 88.93\ts{0.05}      & 93.20\ts{0.42}      & 89.11\ts{0.92}      & 90.35\ts{0.72}      & 94.20\ts{0.47}      & 86.51\ts{0.57}      & 88.97\ts{0.66}      \\
    UNext \cite{valanarasu2022unext}              & 92.47\ts{0.02}      & 85.77\ts{0.05}      & 87.25\ts{0.12}      & 93.17\ts{0.01}      & 86.80\ts{0.41}      & 88.43\ts{0.01}      & 92.42\ts{0.01}      & 85.27\ts{0.30}      & 87.76\ts{0.22}      \\
    CMUNeXt \cite{tang2024cmunext}                & 93.25\ts{0.02}      & 87.15\ts{0.05}      & 88.76\ts{0.05}      & 93.84\ts{0.21}      & 88.39\ts{0.56}      & 90.44\ts{1.29}      & 93.18\ts{0.15}      & 86.39\ts{0.61}      & 88.55\ts{1.29}      \\
    MedT \cite{valanarasu2021medical}             & 91.50\ts{0.06}      & 83.79\ts{0.03}      & 85.35\ts{0.25}      & 92.81\ts{0.46}      & 85.58\ts{0.07}      & 87.28\ts{0.81}      & 90.91\ts{0.34}      & 82.72\ts{0.02}      & 85.41\ts{0.31}      \\
    CMUNet \cite{tang2023cmu}                     & 93.73\ts{0.01}      & 87.91\ts{0.03}      & 89.50\ts{0.10}      & 93.91\ts{0.16}      & 89.10\ts{0.02}      & 89.67\ts{0.36}      & 94.01\ts{0.19}      & 87.17\ts{0.04}      & 90.64\ts{0.49}      \\
    CCT$^{\dagger}$ \cite{ouali2020semi}          & 92.21\ts{0.27}      & 83.94\ts{1.03}      & 88.44\ts{0.12}      & 90.55\ts{1.62}      & 82.76\ts{2.22}      & 89.89\ts{0.95}      & 94.70\ts{1.28}      & 85.95\ts{0.39}      & 88.73\ts{1.18}      \\
    CPS$^{\dagger}$ \cite{chen2021semi}           & 91.74\ts{1.24}      & 82.91\ts{2.38}      & 88.82\ts{0.52}      & 89.35\ts{2.15}      & 80.03\ts{3.70}      & 90.60\ts{0.39}      & 95.06\ts{0.03}      & 87.05\ts{0.71}      & 88.53\ts{0.69}      \\
    CTCT$^{\dagger}$ \cite{luo2022semi}           & 92.51\ts{0.27}      & 84.70\ts{0.46}      & 89.00\ts{0.15}      & 90.69\ts{1.19}      & 83.63\ts{1.66}      & 90.22\ts{0.16}      & 95.13\ts{0.87}      & 86.49\ts{1.02}      & 89.29\ts{0.38}      \\
    ICT$^{\dagger}$ \cite{verma2022interpolation} & 92.89\ts{0.03}      & 85.96\ts{0.16}      & 88.92\ts{0.25}      & 91.73\ts{0.44}      & 85.46\ts{0.76}      & 90.15\ts{0.59}      & 94.77\ts{0.52}      & 87.11\ts{0.54}      & 89.23\ts{0.65}      \\
    R-Drop$^{\dagger}$ \cite{wu2021r}             & 92.27\ts{0.30}      & 84.70\ts{0.64}      & 88.59\ts{0.07}      & 90.10\ts{0.91}      & 83.94\ts{1.42}      & 90.25\ts{0.85}      & \tb{95.36}\ts{0.43} & 86.23\ts{1.10}      & 88.50\ts{1.00}      \\
    URPC$^{\dagger}$ \cite{luo2021efficient}      & 92.96\ts{0.11}      & 85.70\ts{0.58}      & 88.87\ts{0.14}      & 91.90\ts{0.43}      & 85.31\ts{1.93}      & 90.49\ts{0.62}      & 94.65\ts{0.34}      & 86.77\ts{1.00}      & 88.85\ts{0.46}      \\
    MGCC$^{\dagger}$ \cite{tang2023multi}         & 93.79\ts{0.02}      & 88.07\ts{0.00}      & 89.38\ts{0.23}      & 94.06\ts{0.00}      & 89.03\ts{0.20}      & 89.96\ts{0.08}      & 93.98\ts{0.03}      & 87.55\ts{0.19}      & 90.25\ts{0.37}      \\
    AAMS$^{\dagger}$ (Ours)                       & \tb{94.11}\ts{0.00} & \tb{88.48}\ts{0.00} & \tb{90.33}\ts{0.00} & \tb{94.61}\ts{0.01} & \tb{89.33}\ts{0.20} & \tb{91.05}\ts{0.04} & 94.03\ts{0.00}      & \tb{88.07}\ts{0.17} & \tb{90.84}\ts{0.08} \\
\bottomrule
\end{tabular*}
}
\label{tab3}
\end{table*}

We present the segmentation results across AAMS and SOTA methods on the CAMUS dataset in \cref{tab3}. It is evident that our AAMS achieves superior overall performance compared to SOTA methods to different degrees on four structures. Considering the left ventricle endocardium, our AAMS attains the highest DSC and SEN of 94.11\% and 94.61\%. Although R-Drop yields the optimal PRE of 95.36\%, it demonstrates a comparatively low SEN of 90.10\%. In contrast, the MedT produces suboptimal results under this setup with a DSC of 91.50\%. Regarding the epicardium contour, the AAMS achieves the best DSC, SEN, and PRE of 88.48\%, 89.33\%, and 88.07\%. The lowest performance is observed on CPS with a DSC of 82.91\%. For the left atrium, the proposed AAMS presents the optimal DSC, SEN, and PRE of 90.33\%, 91.05\%, and 90.84\%. The MedT consistently presents undesirable results with a DSC of 85.35\%. The performance superiority of our AAMS highlights its generalization, even though the advantages of the ADW scheme are not fully exploited due to the comparable structure size.
 

\section{Conclusion}
\label{7}

In this paper, we develop a novel UNGT dataset to mitigate the dataset gap. The proposed dataset comprises 493 images from 110 patients, featuring an average image resolution of approximately 879 $\times$ 583. Images obtained from different US machines introduce heterogeneity and enhance the robustness and real-world applicability of the dataset. Besides, we introduce a semi-supervised AAMS approach to tackle data limitation and imbalance concurrently. The combination of semi‑supervised learning and adaptive class weighting ensures its effectiveness and transferability for data shortage or highly imbalanced tasks, such as rare‑lesion or under‑represented organ segmentation. As a best‑practice guideline, we recommend deploying dynamic inverse‑frequency weights and maintaining an unlabeled pool that covers all classes to maximize overall performance. We conduct comprehensive segmentation experiments, and the results demonstrate that the developed AAMS shows manifest performance leadership compared with prevalent SOTA methods. Moreover, we perform extensive classification studies on varying SOTA approaches and compare their performance. We believe the first US NGT dataset and novel medical segmenter for tackling data limitation and imbalance can provide valuable insights and benefit the broader medical research community.

The potential approach limitations and further research perspectives are twofold. For one, the introduced UNGT dataset comprises a limited number of images and encounters severely unbalanced structures. To this end, collecting and annotating supplementary images can be accounted for, especially for the underrepresented tube and pancreas structures. With these, the enlarged dataset can comprise a more balanced category distribution. For another, the proposed AAMS trains varying structures with identical iterations. Accordingly, training infrequent or minor structures with additional iterations than the common ones can be beneficial. For example, objects like the tube and pancreas could be trained with more iterations than the liver and stomach. This strategy involves processing all channels in each batch during standard training while dedicating additional iterations to sparse or small categories to emphasize their features. Such an approach could improve the performance of sparse or small objects without compromising the learning of others. The weighting scheme can then be optimized based on such modifications for further performance enhancement.


\section*{Acknowledgements}
This work is supported by Tan Tock Seng Hospital (A-8001334-00-00).



\bibliographystyle{unsrt}
\bibliography{reference.bib}

\biboptions{sort&compress}







\end{sloppypar}
\end{document}